\documentclass[10pt]{article}
\usepackage{amsmath}
\usepackage{amssymb}
\usepackage{graphicx}
\usepackage{epstopdf}
\usepackage[breaklinks,colorlinks=true,linkcolor=blue,citecolor=red,backref=page]{hyperref}

\newcommand{\RR}{{\mathbb R}}
\newcommand{\EE}{{\mathbb E}}

\DeclareMathAlphabet{\itbf}{OML}{cmm}{b}{it}

\newcommand{\bx}{{\itbf x}}
\newcommand{\by}{{\itbf y}}

\newcommand{\bs}{{\itbf s}}

\newcommand{\bk}{\boldsymbol{\kappa}}

\newcommand{\cT}{{\boldsymbol{\cal T}}}
\renewcommand{\hat}{\widehat}

\newcommand{\bzeta}{{\boldsymbol{\zeta}}}

\newcommand{\rc}{\textcolor{black}}

\begin{document}
\title{Multiplexing schemes for optical communication through atmospheric turbulence}
\author{ Liliana Borcea\footnotemark[1] \and Josselin
  Garnier\footnotemark[2] \and Knut S{\o}lna\footnotemark[3] }
\renewcommand{\thefootnote}{\fnsymbol{footnote}}
\footnotetext[1]{Department of Mathematics, University of Michigan,
  Ann Arbor, MI 48109. {\tt borcea@umich.edu}} \footnotetext[2]{Centre
  de Math\'ematiques Appliqu\'ees, Ecole Polytechnique, 91128
  Palaiseau Cedex, France.  {\tt josselin.garnier@polytechnique.edu}}
\footnotetext[3]{Department of Mathematics, University of California
  at Irvine, Irvine, CA 92697. {\tt ksolna@math.uci.edu}}
\maketitle

\begin{abstract}
A central question in free-space optical communications is how to improve the transfer of information between a transmitter and receiver.
The  capacity of the communication channel can be increased by multiplexing of independent modes using either: (1) the MIMO (Multiple-Input-Multiple-Output) approach, where the communication is done with modes obtained from the  singular value decomposition of the transfer matrix from the transmitter array to the receiver array, 
or (2) the OAM (Orbital Angular Momentum) approach, which uses vortex beams that carry angular momenta. 
In both cases, the number of usable  modes is limited by the finite aperture of the transmitter and receiver, and  the effect of 
the turbulent atmosphere. The goal of this paper is twofold: First,  we show that the MIMO and OAM  multiplexing schemes are closely related.
Specifically, in the case of circular apertures, the usable singular modes of the transfer matrix are essentially the same as the commonly used 
Laguerre-Gauss vortex beams, provided these have a special radius that depends on the wavelength, the distance from the transmitter to the receiver and the ratio of the radii of their apertures. Second, we study the effect of atmospheric turbulence on the communication modes using the 
phase screen method put in the mathematical framework of beam propagation in random media. 
\end{abstract}

\section{Introduction}
In free-space optical communications, one seeks to transfer information between a transmitter array and a receiver array using laser beams. 
It is an important technology for line-of-sight communication between moving locations (e.g. in satellite communication), or for settings where fiber-based systems do not exist. 
A central question is how to increase the capacity of the communication channel via multiplexing independent sub-channels called modes. Typically, these are defined as special orthogonal solutions of the Helmholtz equation in homogeneous transmission media. 
Orbital Angular Momentum (OAM) beams \cite{gibson04,Barnett} are such solutions, also known as vortex beams \cite[Chapter 2]{gbur16}, because they exhibit a vortex on the axis of the beam, where the intensity is zero and the phase is not defined. Popular examples of OAM beams are: (1) Bessel beams, which 
have the desirable nondiffractive property, but cannot be realized in practice as they carry infinite energy \cite{bouchal03}. 
Therefore, they are approximated via some 
truncation strategy to obtain, for example, the Bessel-Gauss beams \cite{Gori}. (2) Hermite-Gauss and Laguerre-Gauss beams \cite[Section 16]{Siegman}, which are solutions of the paraxial approximation of the Helmholtz equation in rectangular and cylindrical coordinates, respectively. 
The Laguerre-Gauss beams are of special interest because they are easily realizable in practice \cite{Anguita,willner17}.

OAM beams have received much attention because by changing their azimuthal angle $\theta$ dependency $\exp (i j \theta)$, where the integer $j $ is the so-called topological charge, one can create in theory an infinite number of modes. 
However,  the number of usable modes is limited in practice by  the finite size of the apertures of the transmitter and receiver, the interaction of the beam with the atmosphere and noise.  If these factors are not taken into account, then the modes are no longer orthogonal at the receiver end, meaning that there is channel cross-talk and loss of information \cite{Anguita,doster}.

The finite aperture size can be accounted for by using a Singular Value Decomposition (SVD) based Multiple-Input-Multiple-Output (MIMO) multiplexing approach, where the modes are singular vectors of the transfer matrix \cite{Miller}. The transmission modes are the right singular vectors and the received modes are the left singular vectors. The  SVD gives the optimal useable communication modes, corresponding to the significant singular values. Ideally, these modes should be determined by  measuring the transfer matrix  and then carrying out its  SVD decomposition. However, the atmosphere changes in time and repeated measurements of the transfer matrix may be difficult to make in practice. This has motivated studies like \cite{Anguita,doster,rodenburg}, which seek to quantify with numerical simulations the effect of atmospheric turbulence on modes calculated for a proxy transfer matrix in a synthetic homogeneous medium.

A much debated issue has been the advantage of using OAM beams versus MIMO multiplexing \cite{Chen,Edfors,Morabito}. Here we show that 
in fact the two approaches are closely related. We consider a paraxial beam propagation model and assume that we can approximate the transmission and receiver arrays by continuous disk shaped apertures. Then, the SVD of the proxy transfer matrix  has an explicit solution 
in terms of the circular prolate spheroidal functions \cite{slepian4,lederman}. If we let $a_R$ and $a_T$ be the radii of the receiver and transmitter
arrays, and denote by $\lambda$ the wavelength and  $L$ the transmission distance, the matrix has 
\begin{equation}
\label{eq:Nmodes}
N = \frac{\pi a^2_R}{[\lambda L/(2a_T)]^2}
\end{equation}
singular values that are very close to one, and the remaining ones plunge rapidly toward zero. A large singular value means that 
the mode carries large power within the receiver aperture and is thus less affected by noise.
Physically, the number $N$ can be interpreted as the number of focal spots of linear size $\lambda L/(2a_T)$ that fit in the receiver aperture area $\pi a_R^2$. The interesting regime for communications corresponds  to a large $N$, so that one can multiplex the many $N$ useable modes.  

It has been observed in \cite{shapiro05,RodenburgPhD} that in the case of soft apertures, with Gaussian apodization, the convenient Laguerre-Gauss beams are the modes given by the SVD. The similarity of some Laguerre-Gauss beam modes and the circular prolate spheroidal functions was also noticed in \cite{RodenburgPhD,Miller}. Using the theory of circular prolate spheroidal functions \cite{slepian4}, we show that in fact the 
Laguerre-Gauss beam modes are the significant modes given by the SVD even for hard aperture thresholding, as long as their radius is carefully calibrated  in terms of the wavelength $\lambda$, the propagation distance $L$, and the ratio $a_T/a_R$ as prescribed by the forthcoming formula (\ref{def:radiusLG}).

Some recent studies \cite{doster} have found that Bessel-Gauss OAM beams are more robust to turbulence effects than the Laguerre-Gauss beams.
Here we show that, provided the  Laguerre-Gauss beams are calibrated as stated above, the opposite is true. We study the effect of turbulence using the mathematical theory of paraxial wave propagation in random media with statistics corresponding to Kolmogorov type turbulence. We use the  theory to put numerical phase screen simulation results  \cite{paterson,tyler} in a mathematical framework and to clarify the 
link  between the phase screen parameters and the turbulence model. 

The paper is organized as follows.
In Section \ref{sec:1} we consider laser beam propagation in homogeneous free space and study different candidates
to multiplexing schemes. In particular, we identify the leading singular vectors  of the transfer matrix that are used in the MIMO approach and show that they are related to Gauss-Laguerre modes.
In Section \ref{sec:2} we consider laser beam propagation through the turbulent atmosphere 
and give the It\^o-Schr\"odinger mathematical model  which characterizes the statistics of the transmitted beams.
In Section \ref{sec:3} we quantify the robustness of different multiplexing schemes with respect to the turbulent atmosphere.
We conclude with a brief summary in section \ref{sec:summary}.

\section{Homogeneous paraxial wave equation}
\label{sec:1}
In this section we describe classical beams that exhibit orthogonality when they propagate through a homogeneous medium.
They are approximate solutions of the Helmholtz  equation of the form  $u(r,\theta,z) \exp(i k z)$, with $u(r,\theta,z)$
satisfying the  paraxial wave equation \cite{Siegman},
\begin{equation}
2 i k \partial_z u(r,\theta,z) + \Delta_\perp u(r,\theta,z)=0,
\label{eq:paraxu}
\end{equation}
where $k = 2 \pi/\lambda$ is the wavenumber.
Because we assume circular apertures of the transmitter and receiver arrays, we use the cylindrical coordinates   $(r,\theta,z)$, with $z$ measured along the axis of the beam, radius $r$ and azimuth $\theta$. The operator $\Delta_\perp= {r}^{-1} \partial_r \big( r \partial_r \cdot \big) + r^{-2} \partial_\theta^2$  is the transverse Laplacian.
We assume throughout an input beam profile $u(r,\theta,z=0)$ with slow variation with respect to $\lambda$, so that the paraxial approximation is valid.

\subsection{Bessel-Gauss beams}
Let $\beta>0$ and $r_o>0$ be such that $k  \gg \beta$ and $ k r_o \gg 1$.
For any integer $j$, 
the input profile of the $j$-th Bessel-Gauss beam is \cite[Section 12.1]{gbur16}
\begin{equation}
\label{eq:inputBG}
u_j^{\rm BG}(r,\theta,z=0) =  
J_{j}\big(\beta r\big)
\exp \Big( - \frac{r^2}{r_o^2}\Big)\exp \big( i j  \theta\big) ,
\end{equation}
where $J_j$ is the Bessel function of the first kind.
After propagation over a distance $z$ in the homogeneous medium, the output profile is
\begin{align}
\nonumber
u_j^{\rm BG}(r,\theta,z) =& \frac{r_o}{r_z}
J_{j}\Big( \frac{\beta r}{1+i\frac{z}{z_{\rm R}}}\Big)
\exp\left[ \Big(r^2+ \frac{\beta^2z^2}{k^2}\Big) \Big( - \frac{1}{r_z^2}+ i \frac{k}{2R_z}\Big)\right]
 \\
&\times \exp\left[ i j \theta -i \frac{\beta^2}{2k} z - i \, {\rm atan}\Big(\frac{z}{z_{\rm R}}\Big)\right] ,
\label{eq:outputBG}
\end{align}
where $z_{\rm R} = kr_o^2/2$ is the Rayleigh length,
$r_z$, resp. $R_z$, is the radius  of a standard Gaussian beam at distance $z$, resp.  the radius of curvature of the wavefront:
\begin{equation}
r_z=r_o \Big(1+ \frac{z^2}{z_{\rm R}^2}\Big)^{\frac{1}{2}},\quad R_z=z \Big(1+ \frac{z_{\rm R}^2}{z^2}\Big)  .
\label{eq:rz}
\end{equation}
If $r_o\to+\infty$ then \eqref{eq:outputBG} tends to
$$
u_j(r,\theta,z) =
J_{j}(\beta r)
\exp \Big( i j \theta -i \frac{\beta^2}{2k}  z\Big)  ,
$$
the ideal $j$-th Bessel beam  \cite[Section 12.1]{gbur16} which is diffraction-free, but cannot be realized in practice as it has infinite energy ($L^2$-norm).

%
%

\subsection{Laguerre-Gauss beams}
Let $r_o>0$ be such that $ k r_o \gg 1$
and $p,j$ be integers with $p \geq 0$.
The input profile of the $(p,j)$-th Laguerre-Gauss mode is \cite[Section 2.2]{gbur16}
\begin{align}
\nonumber
 u^{\rm LG}_{p,j}(r,\theta,z=0) =& 
\sqrt{\frac{2p!}{ 
\pi (|j|+p)!}}
\Big(\frac{\sqrt{2}r}{r_o}\Big)^{|j|}  L_p^{|j|}\Big(\frac{2r^2}{r_o^2}\Big) \\
&\times \exp \Big( - \frac{r^2}{r_o^2}  
+ i j \theta\Big),
\label{eq:LG0}
\end{align}
where $L_p^j$ is the  generalized Laguerre polynomial $L_p^j(s) =\frac{ e^s s^{-j}}{p!} \frac{d^p}{ds^p} (e^{-s} s^{p+j})$.
The input Laguerre-Gauss profiles are not compactly supported. However, their essential supports 
are disks with radii of the order of $ r_o$ for the low-order modes, and of the order $\sqrt{|j|} r_o$ for high mode indexes $|j|$ \cite{padgett95}.

After propagation over a distance $z$ in the homogeneous medium, the Laguerre-Gauss beam profiles are
\begin{align}
\nonumber
u^{\rm LG}_{p,j}(r,\theta,z) = &
\sqrt{\frac{2p!}{ 
\pi (|j|+p)!}}
\frac{r_o}{r_z}
\Big(\frac{\sqrt{2}r}{r_z}\Big)^{|j|} L_p^{|j|}\Big(\frac{2r^2}{r_z^2}\Big)\\
\nonumber
&\times \exp \left[ r^2 \Big(- \frac{1}{r_z^2} +   \frac{i k }{2 R_z} \Big)\right] \\
&\times \exp \left[ i j \theta  - i ( |j|+2p+1 ) {\rm atan}\Big(\frac{z}{z_{\rm R}}\Big) \right].
\end{align}
The beams widen due to diffraction, as modeled by the beam radius $r_z$ and the radius $R_z$ of curvature of the wavefront
defined in \eqref{eq:rz}.

%

\subsection{SVD based MIMO multiplexing}
\label{sec:SVD}
Bessel-Gauss and Laguerre-Gauss beams are two of the many examples of orthogonal modes that carry an angular momentum i.e., a phase of the form $\exp(i j \theta)$ which  is kept invariant during the propagation. In theory, for transmission through the homogeneous medium, and for infinite transmitter and receiver apertures, the countably infinite family of such orthogonal modes could be used to obtain an indefinite increase in the capacity of the communication channel. In reality, this cannot be achieved due to the finite transmitter and receiver apertures and heterogeneity in the transmission medium.  We describe here the limitations imposed by the finite apertures and postpone until section \ref{sec:2} the discussion of the effect of a turbulent transmission medium. 

A systematic approach for describing which beams are most appropriate for communication between a transmitter and receiver array is given by the SVD of the transfer matrix $\cT$. Assuming that the transmitter array has $n_T$ elements and the receiver array has $n_R$ elements,  this is an $n_R \times n_T$  matrix with complex entries ${\cal T}_{t,r}$ corresponding to the complex wave amplitude at the $r-$th receiver, due to a unit input at the $t-$th transmitter. The matrix $\cT$ can be computed by solving the wave equation in the homogeneous medium. Its  right singular vectors   are the orthonormal input profiles  that can be used in multiplexing at the transmitter array. The left singular vectors  form  the orthonormal basis that can be used for demultiplexing at the receiver array. 

\subsection{SVD in the continuum approximation}
\label{sec:slepSVD}
If the transmitters and receivers are closely spaced in the arrays with radii $a_T$ and $a_R$, in the sense that their linear size is smaller than  the Rayleigh resolution limit $\lambda L/ [2 \max\{a_T,a_R\}]$,  we can approximate the arrays by the continuous apertures
\begin{align*}
{\cal A}_{T,R} = \{ \bx = (r \cos \theta, r \sin \theta), ~ 0 \le r \le a_{T,R}, ~ \theta \in [0,2\pi]\}.
\end{align*}
In this continuous setting, the transfer matrix becomes the linear  integral operator  $\cT:L^2({\cal A}_T) \mapsto L^2({\cal A}_R)$, 
\begin{equation}
u(\bx,L) = \cT u_o(\bx) = \int_{{\cal A}_T}  u_o(\bx')  G\left((\bx,L),(\bx',0)\right)d \bx', 
\label{eq:TransfM}
\end{equation}
for $\bx \in {\cal A}_R$,
where $u_o(\bx) = u(\bx,0)$ 
is the input beam profile at the transmitter array and the kernel is  the Green's function of the paraxial equation (\ref{eq:paraxu}),
\begin{equation}
G\left((\bx,L),(\bx',0)\right) = \frac{k}{2 i \pi L} \exp \Big( i \frac{k |\bx-\bx'|^2}{2 L} \Big).
\label{eq:GF}
\end{equation}

The right singular functions of $\cT$, which define the transmission modes in the MIMO multiplexing, are of the form
\begin{equation}
u_o(\bx) =  \exp\Big( - i \frac{k|\bx|^2}{2L}\Big) v \Big( \frac{\bx}{a_T} \Big), \quad \bx \in {\cal A}_T,
\label{eq:RSVD}
\end{equation}
where $v(\bs)$ are the right singular functions of the linear integral operator ${\cal S}:L^2(B({\bf 0},1)) \mapsto L^2(B({\bf 0},1))$ defined by
\begin{equation}
\label{eq:opslepian}
{\cal S} v(\bs) =  \int_{B({\bf 0},1)} v(\bs') \exp\big( - i C \bs \cdot\bs'\big) d\bs', \qquad \bs \in B({\bf 0},1),
\end{equation}
with  $B({\bf 0},1)$ the unit disk centered at the origin ${\bf 0}$ of the cross-range plane and 
\begin{equation}
C=\frac{k a_T a_R }{L} . \label{eq:defC}
\end{equation}
The operator  ${\cal S}$ was studied by Slepian \cite{slepian4}. Its singular functions $v(\bs)$ are the generalized prolate spheroidal functions, 
its first 
$N = C^2/\pi$ singular values (recall \eqref{eq:Nmodes})  are close to $1/C$, and the remaining ones plunge rapidly to zero.

We are interested in the case $C \gg 1$, where there are $N \gg 1$ transmission modes of the form (\ref{eq:RSVD}) available for multiplexing. 
For such $C$, it follows from \cite[Eq.~(67)]{slepian4} that the leading singular functions of ${\cal S}$ behave like scaled Gauss-Laguerre functions \begin{equation}
v_{p,j}(\bs) = \exp\Big( - \frac{C |\bs|^2}{2}\Big) L_p^{|j|} (C |\bs|^2) (\sqrt{C} |\bs|)^{|j|} e^{i j {\rm arg}(\bs)}   ,
\label{eq:vpj}
\end{equation}
for integers $p,j$, with $p \ge 0$. Thus, we conclude from \eqref{eq:RSVD}  that the transmission modes are of the form (\ref{eq:LG0}) up to multiplicative constants,
\begin{equation}
\label{express:eigenslepian}
u_{p,j}(\bx,z=0) =  \exp\Big( - i \frac{k|\bx|^2}{2L} - \frac{|\bx|^2}{r_o^2}\Big) L_p^j \Big( \frac{2|\bx|^2}{r_o^2}\Big) |\bx|^j 
e^{i j {\rm arg}(\bx)}  ,
\end{equation}
for the special radius
\begin{equation}
\label{def:radiusLG}
r_o= \frac{\sqrt{2L a_T}}{\sqrt{k a_R}}.
\end{equation}
They correspond to the following leading singular values of $\cT$   \cite[Eq.~(93)]{slepian4},
\begin{equation}
\label{def:lambdapj}
\mathfrak{S}_{p,j} = 1 - \mathfrak{S}_{p,j}' \big[1 +O(C^{-1})\big] , 
\end{equation}
with
\begin{equation}
\label{def:lambdapj2}
   \mathfrak{S}_{p,j}'=\frac{\pi 2^{2|j|+4p+3} C^{|j|+2p+1} e^{-2C}}{p! (p+|j|)!} . 
\end{equation}

Note that the quadratic phase in the first factor in \eqref{express:eigenslepian} makes the beam focus at distance (beam waist) $L/(1+a_R^2/a^2_T)$. 
The beam then diffracts from there to the receiver array, to get an output profile that is similar to the emitted one, but rescaled 
by the radius $a_R$.

\begin{figure}[t]
\begin{center}
\begin{tabular}{c}
\includegraphics[width=4.35cm]{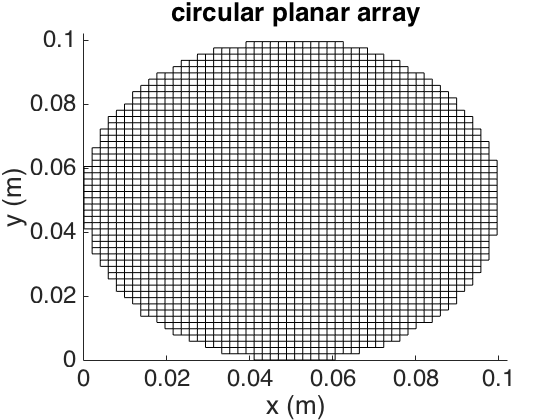} 
\includegraphics[width=4.35cm]{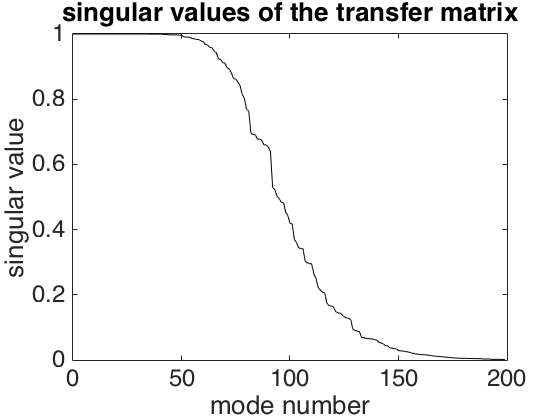} 
\end{tabular}
\end{center}
\vspace{-0.2in}\caption{Left: Circular planar array with $n=2063$ square elements with size $2$mm in the disk shaped aperture ${\cal A}$ of radius $a=5$cm.
Right: The first  $200$ singular values of the transfer matrix for transmission distance $L=1$km and wavelength $\lambda = 850$nm.}
\label{fig:circularplanararray}
\end{figure}

\subsection{Illustration}
\label{sec:illustrate}
We consider throughout a practical setup for free-space optical communication with laser beams \cite{doster} at wavelength $\lambda = 850$nm,  using  transmitter and receiver arrays with the same circular aperture ${\cal A}$, of radius $a = a_T = a_R = 5$cm. The arrays have $n = n_T = n_R = 2063$ square elements with side length $2$mm (see left plot in Figure~\ref{fig:circularplanararray}). The transmission distance is $L = 1$km. 
Note that the aperture is not centered at the origin $(0,0)$, but at 
\rc{$(5,5)$cm}. All the beam axes are shifted to pass through this center.

The singular value decomposition described in section \ref{sec:slepSVD}  is relevant here, because the Rayleigh resolution limit $\lambda_o L /(2a) = 8.5$mm is larger than the $2$mm size of the elements of the arrays. 
After computing the $n \times n$ transfer matrix in the homogeneous medium and carrying out its SVD, we find the singular values displayed in the right plot in Figure \ref{fig:circularplanararray}.
There are approximately $100$ large ones, which is very close to the theoretical estimate (\ref{eq:Nmodes}) of the essential rank of the integral operator (\ref{eq:TransfM}) that predicts $N=109$. 
The right singular vectors give the orthonormal input profiles to be used in the MIMO multiplexing. 
The left singular vectors give the basis on which we project the wave at the receiver array, for demultiplexing.
In our illustration the transmitter and receiver arrays are identical, so the transfer matrix is complex symmetric and the right and left singular vectors are the same. We ensure that this is the case in the computations by using the symmetric (Takagi) SVD. 

\begin{figure}[th]
 \begin{center}
  \begin{tabular}{c}
\includegraphics[width=2.85cm]{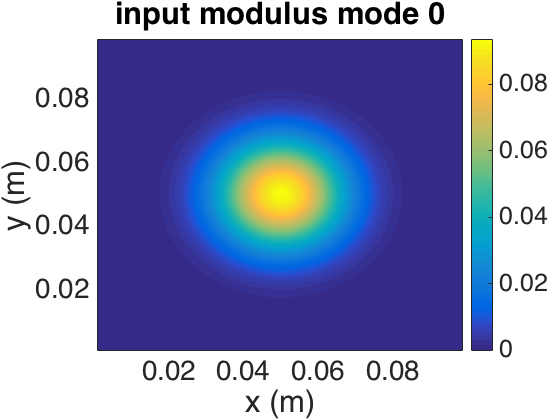}  
\includegraphics[width=2.85cm]{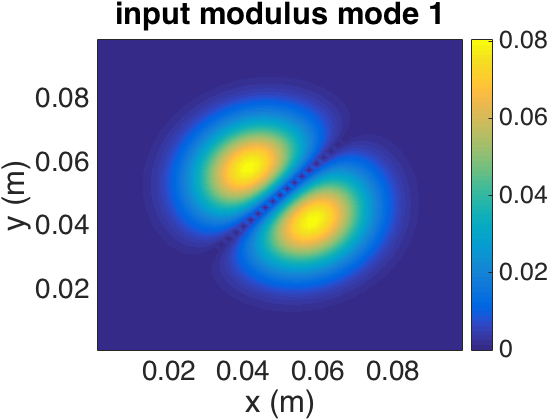}  
\includegraphics[width=2.85cm]{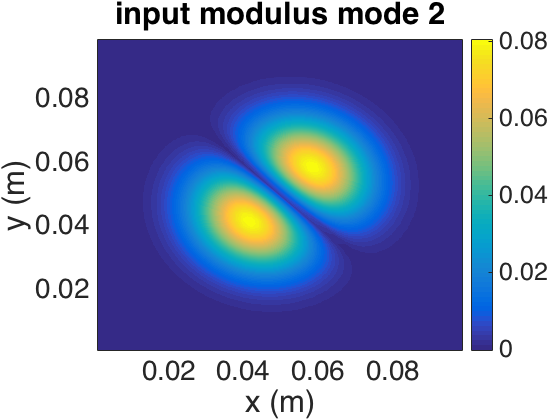}\\  
\includegraphics[width=2.85cm]{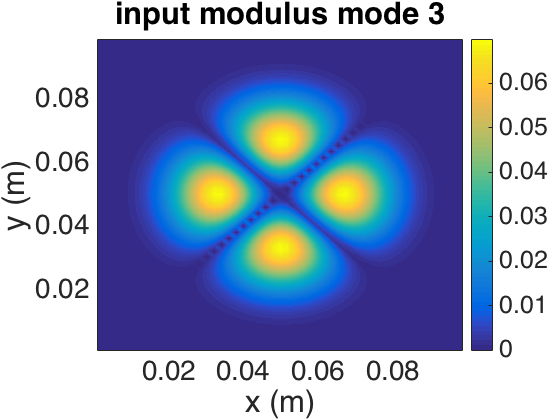}  
\includegraphics[width=2.85cm]{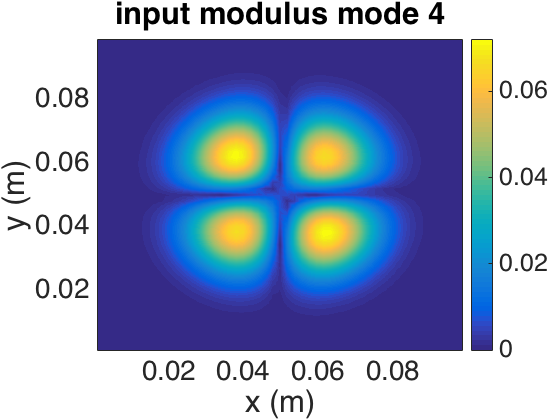}  
\includegraphics[width=2.85cm]{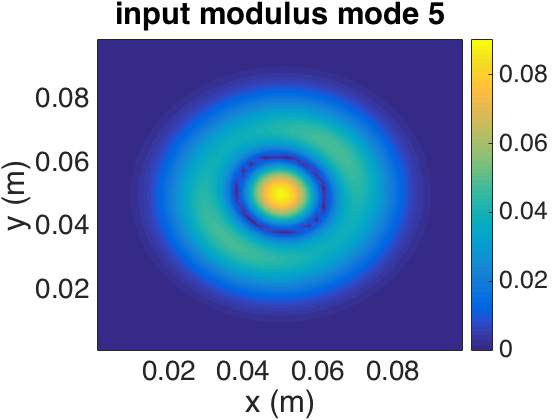}  \\
\includegraphics[width=2.85cm]{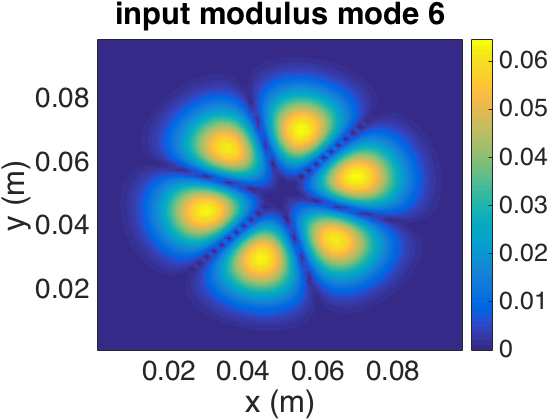}  
\includegraphics[width=2.85cm]{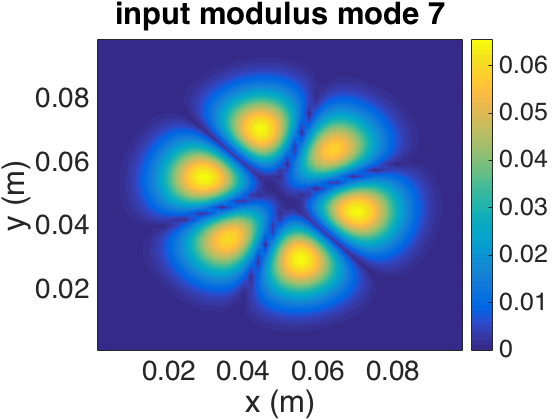}  
\includegraphics[width=2.85cm]{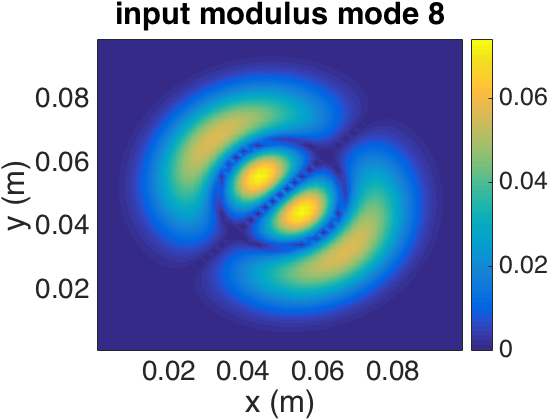}  \\
\includegraphics[width=2.85cm]{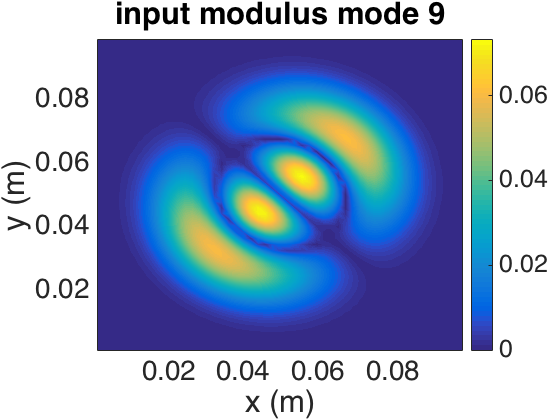} 
\includegraphics[width=2.85cm]{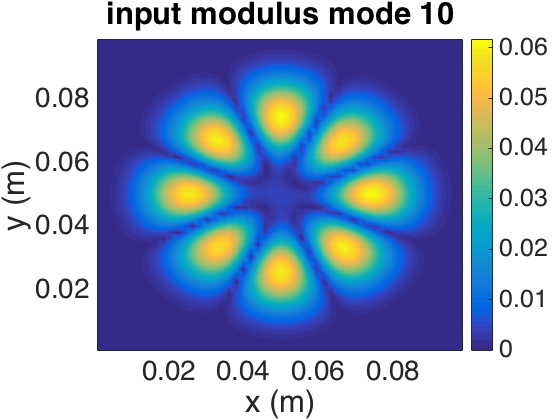}  
\includegraphics[width=2.85cm]{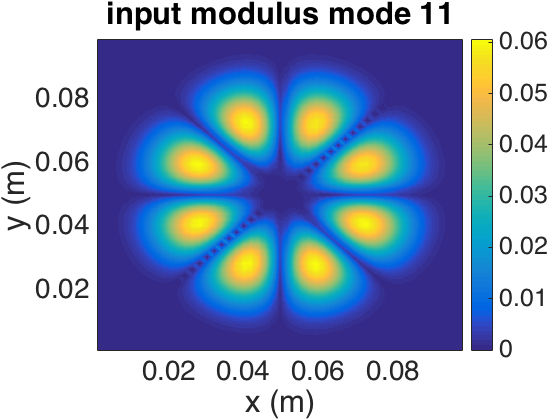}  
\end{tabular}
\end{center}
\vspace{-0.2in}
\caption{Moduli of the mode profiles. The axes are in meters. }
\label{fig:eigen}
\end{figure}

\begin{figure}[h!]
 \begin{center}
  \begin{tabular}{c}
\includegraphics[width=2.85cm]{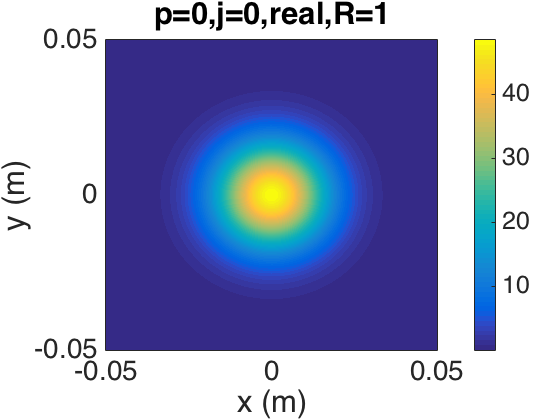}  
\includegraphics[width=2.85cm]{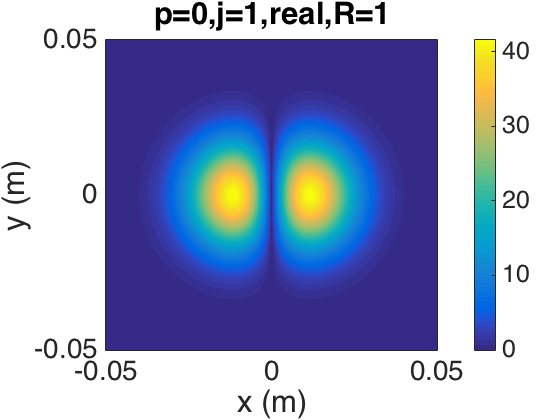}  
\includegraphics[width=2.85cm]{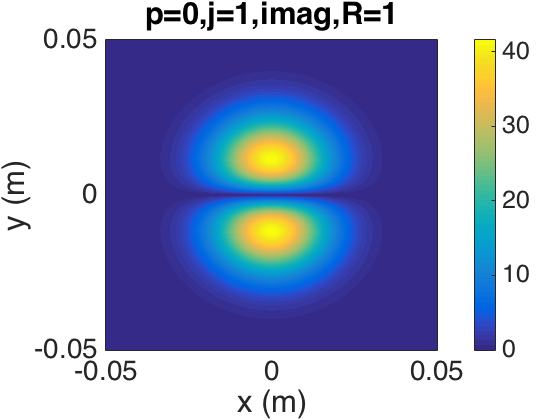}\\  
\includegraphics[width=2.85cm]{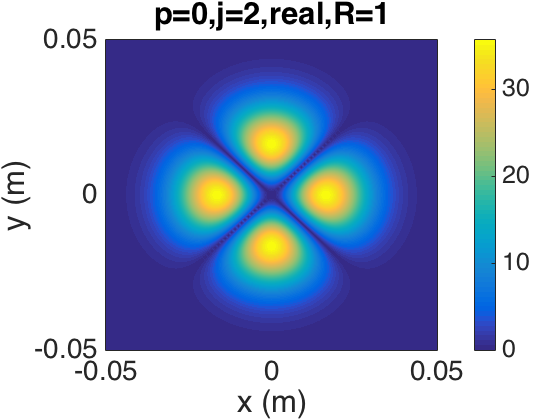}  
\includegraphics[width=2.85cm]{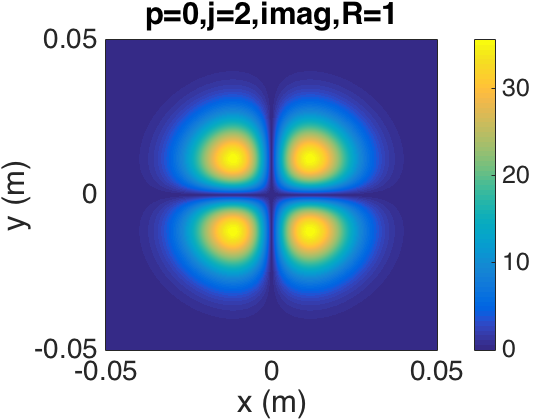}  
\includegraphics[width=2.85cm]{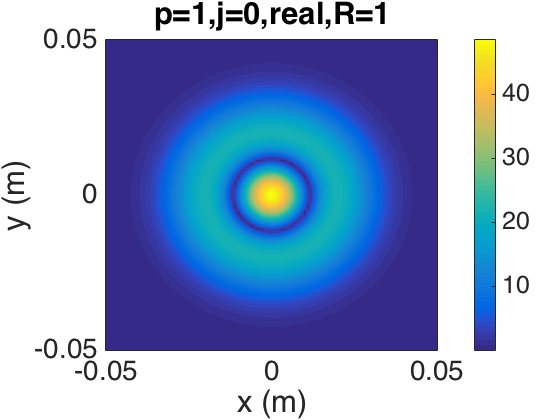}  \\
\includegraphics[width=2.85cm]{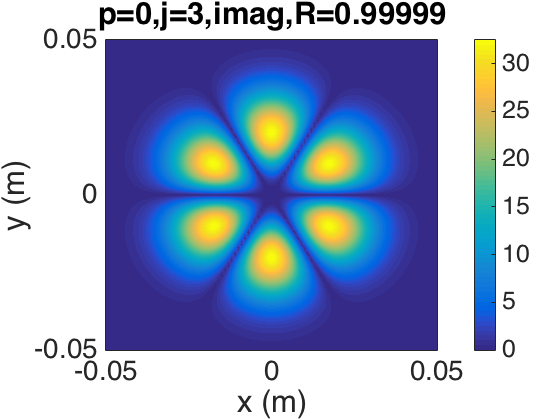}  
\includegraphics[width=2.85cm]{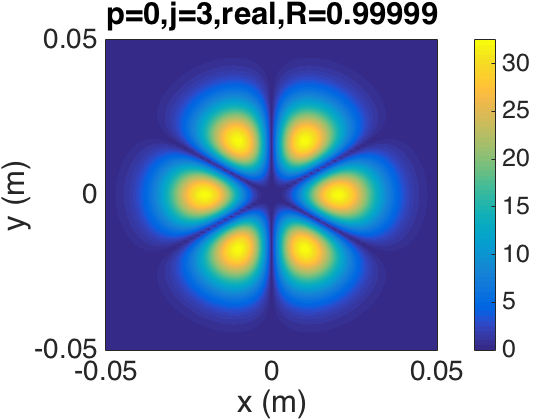}  
\includegraphics[width=2.85cm]{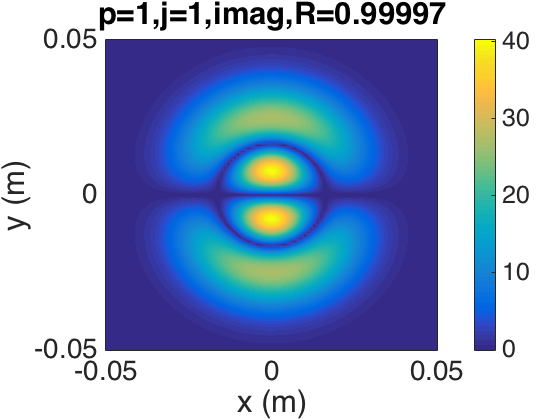}  \\
\includegraphics[width=2.85cm]{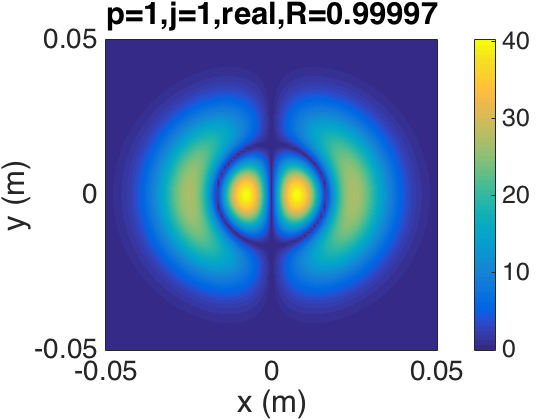} 
\includegraphics[width=2.85cm]{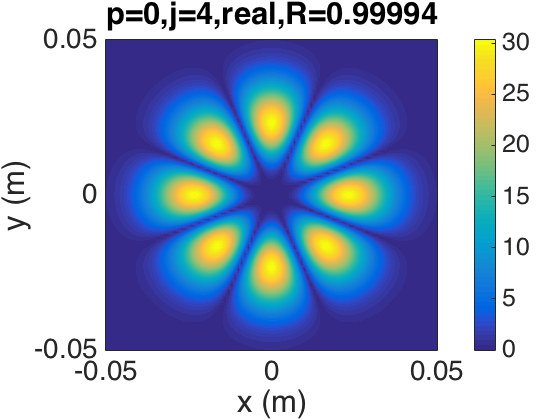}  
\includegraphics[width=2.85cm]{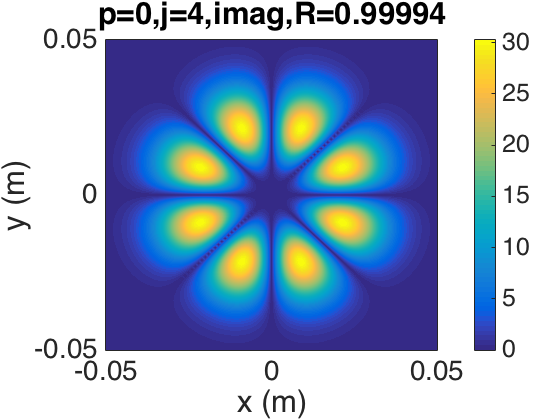}  
\end{tabular}
\end{center}
\vspace{-0.2in}\caption{Absolute values  of the real and imaginary parts of the Laguerre-Gauss modes with radius $r_o=1.64$cm.
The profiles are sorted according to the power fraction $R$ in the transmitter aperture.
The axes are in meters.}
\label{fig:lgRI}
\end{figure}

The mode profiles are stepwise constant functions  on the elements of the array, with values given by the singular vectors.  We plot in Figure \ref{fig:eigen} the moduli of the first $12$ modes.  For comparison, we also display in Figure \ref{fig:lgRI} the absolute values of the real and imaginary parts of the Laguerre-Gauss modes with initial radius $r_o = 1.64$cm calculated from \eqref{def:radiusLG}. These modes are sorted according  to their power fraction  in ${\cal A}$. As expected from \eqref{express:eigenslepian}, the plots in Figures \ref{fig:eigen} and  \ref{fig:lgRI} are basically the same, aside from a rotation by the angle $\pi/4$, which is irrelevant because ${\cal A}$ is rotation invariant.

Eqs.~(\ref{def:lambdapj}-\ref{def:lambdapj2})  estimate the leading singular values, with corrections $\mathfrak{S}'_{p,j}$  displayed in Table \ref{table:lambda}. The sorting displayed in this table is the same as the sorting based on the power fraction 
in ${\cal A}$, used  in Figure \ref{fig:lgRI}. 
\begin{table}[h]
\begin{tabular}{|c||c|c|c|c|c|c|c||}
\hline
$(p,j)$ &  
$(0,0)$ &  
$(0,1)$ & 
$(0,2)$ & 
$(1,0)$ & 
$(0,3)$ & 
$(1,1)$ & 
$(0,4)$ \! \\
\hline
\! $\mathfrak{S}_{p,j}'$ &
 $4\cdot 10^{-14}$ & 
$3\cdot 10^{-12}$ &
$10^{-10}$ &
$2\cdot 10^{-10}$ &
$3\cdot 10^{-9}$ &
$8\cdot 10^{-9}$ &
$5 \cdot 10^{-8}$ \\
\hline
\end{tabular}
\caption{Values of $\mathfrak{S}_{p,j}'$ sorted in increasing order.}
\label{table:lambda}
\end{table}

\subsection{Discussion}
The results in section  \ref{sec:SVD} reconcile the MIMO approach, where the singular vectors of the transfer matrix are used for multiplexing,
and the OAM approach, in which Laguerre-Gauss beams  are used as transmission modes \cite[section 6.2]{gbur16}. 
The useful modes correspond to the $N$ large singular values or, equivalently, the Laguerre-Gauss beams with large power within the aperture, 
and they coincide for the two multiplexing approaches, 
provided that the initial radius $r_o$ is chosen as in \eqref{def:radiusLG}. Note that for this radius,  the Rayleigh length satisfies
$
z_R = {k r_o^2}/{2} = L {a_T}/{a_R},
$
so the larger the transmitter aperture, the smaller the diffraction effect.

At large mode numbers, corresponding to negligible power within the aperture, the Laguerre-Gauss modes  and the singular vectors differ, because the latter are compactly supported in the aperture and the former extend outside the aperture. Obviously, such modes are not useful for transmitting information.

\section{Communication through turbulence}
\label{sec:2}

The results in section \ref{sec:1} show that in the ideal case of transmission through a homogeneous medium, the leading $N$ modes
determined from the SVD decomposition of the transfer matrix should be used as transmission modes. Moreover, 
these modes are  the Laguerre-Gauss beams with initial radius (\ref{def:radiusLG}) in the case of dense planar circular arrays and $N \gg 1$.  
We now seek to quantify how such a multiplexing scheme degrades in  a turbulent transmission medium.     
 
 It was shown in  \cite{doster} via numerical simulations, which do not account for finite transmitter and receiver apertures, that Bessel-Gauss beams outperform Laguerre-Gauss beams in channel efficiency through a turbulent medium.  The earlier paper \cite{paterson} studied \rc{the probability of detection of the angular momentum when a Laguerre-Gauss mode is transmitted  through a turbulent medium by using a formal propagation model}.
A similar approach  was used in \cite{tyler},  where the role of  the turbulence strength measured in terms of the 
Fried parameter (relative to the aperture)  was discussed.   
Further insight using this framework was provided in \cite{rodenburg},  where diffractive effects for relatively small aperture 
were discussed both from analytic and experimental viewpoints. 

Here we present a framework where the role of the turbulence is taken into account in a rigorous fashion, and connect it to the  
phase screen model for numerical wave propagation. We also derive explicit formulas for the cross-correlations of the 
wave field in a specific scaling regime, corresponding to weak diffraction. These formulas give good predictions of 
the performance of MIMO and OAM multiplexing schemes, even for moderate diffraction, as explained in section \ref{sec:3}.

\subsection{Random paraxial wave equation and phase screen}
\label{sec:2.1}
Beam propagation  through a turbulent medium can be described mathematically by 
the random paraxial wave equation
\begin{align}
2 i k \partial_z u(\bx,z) + \Delta_\perp u(\bx,z) + {k^2}V(\bx,z) u(\bx,z)&=0, \label{eq:para1}
\end{align}
for $\bx \in \RR^2$ and $z > 0$, with initial condition
\begin{align}
u(\bx,0)&=u_o(\bx),\label{eq:para2}
\end{align}
where $V(\bx,z)$ is a random potential.
We are interested in a phase screen method (i.e., a split-step Fourier method with grid step $\ell_z$) for solving this equation.
This amounts to assuming that the random potential is stepwise constant in $z$ over intervals with length $\ell_z$,
\begin{equation}
\label{eq:modelV}
V(\bx,z) = \sum_{n\geq 0} {\bf 1}_{[n\ell_z,(n+1)\ell_z)} (z) V_n(\bx).
\end{equation}
Here $V_n(\bx)$ are i.i.d. copies of a stationary two-dimensional Gaussian, zero-mean random field 
with covariance function 
\begin{equation}
\EE [ V_n(\bx) V_n(\bx') ] = {\cal R}(\bx -\bx') .
\end{equation}

We assume  isotropic statistics, with covariance given by  the Mat\'ern model
${\cal R}(\bx ) = {\cal R}_\nu(|\bx|)$, 
\begin{equation}
{\cal R}_\nu(r)=
\frac{\sigma^2}{\Gamma(\nu) 2^{\nu-1}} \left( \frac{2 \sqrt{\nu} r}{\ell_c}\right)^\nu
K_\nu \left( \frac{2 \sqrt{\nu} r}{\ell_c}\right) ,
\label{def:materncov}
\end{equation}
and power spectral density 
\begin{equation}
\hat{\cal R}(\bk) = \int_{\RR^2} {\cal R}(\bx) e^{- i \bk\cdot\bx}d\bx = \sigma^2
\frac{2^{2\nu+2} \pi  \nu^{\nu+1}}{  \ell_c^{2\nu}}
\left( \frac{4\nu}{ \ell_c^2}  +|\bk|^2\right)^{-\nu-1} . \label{def:mathernPSD}
\end{equation}
This model depends  on three hyperparameters $\sigma^2,\ell_c,\nu$, and $K_\nu$ is the modified  Bessel function of second kind. 
The hyperparameter $\nu \in [1/2,\infty)$ characterizes the smoothness of the process
(the realizations are $\nu'$-H\"older continuous, for any $\nu' < \nu$),
$\sigma^2 = {\cal R}({\bf 0})$ is the variance  and $\ell_c$ is the correlation radius.
In the limit $\nu \to \infty$, we obtain from \eqref{def:materncov} the smooth Gaussian covariance model
$
{\cal R}(\bx ) = \sigma^2 \exp \big(- {|\bx |^2}/{\ell_c^2}\big),
$
whereas the other extreme $\nu =1/2$  gives the rough exponential covariance model
$
{\cal R}(\bx) = \sigma^2 \exp \big(- {\sqrt{2} |\bx |}/{\ell_c}\big)  .
$ 

\begin{figure}[th]
\begin{center}
\begin{tabular}{c}
\includegraphics[width=4.35cm]{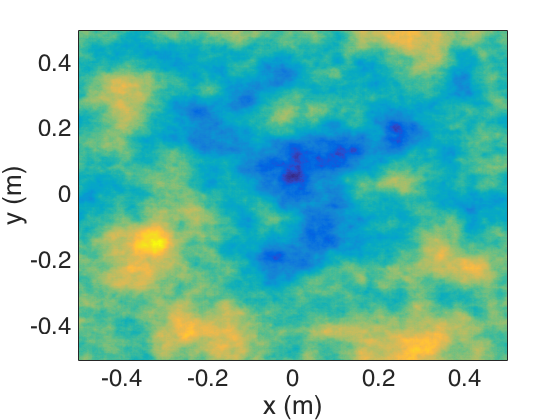} 
\includegraphics[width=4.35cm]{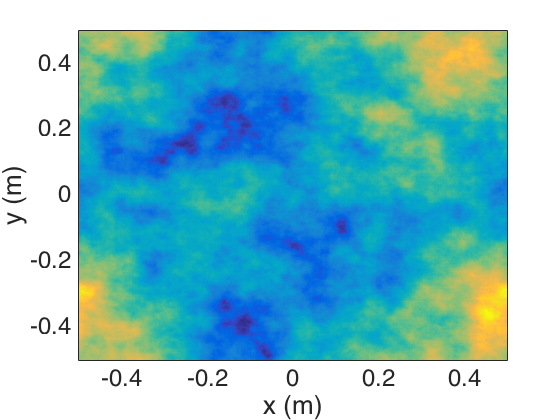} 
\end{tabular}
\end{center}
\vspace{-0.2in}\caption{Two independent phase screens, with $\ell_c=20~$cm ($L_0=37$~cm). The coordinates $x$ and $y$ in the cross-range plane are  in meters.}
\label{fig:phasescreen}
\end{figure}

We consider henceforth $\nu=5/6$, so that \eqref{def:materncov} gives a Kolmogorov-type model with outer scale proportional to $\ell_c$ and inner scale equal to zero.
More explicitly, we recover from \eqref{def:mathernPSD} the standard Kolmogorov  model 
$$
\hat{\cal R}(\bk) = 0.033 C_n^2 \bigg( \frac{1}{L_0^2}  +|\bk|^2\bigg)^{-11/6} ,
$$
if we set $\nu=5/6$, $\ell_c= 2\sqrt{\nu}L_0$, and $\sigma^2 = 0.033 C_n^2 L_0^{2\nu} /  (4\pi\nu)$.
With this parameterization, $L_0$ is the outer scale of turbulence and $C_n^2$ is the  turbulence strength.
In Figure \ref{fig:phasescreen} we plot two realizations of a phase screen obtained with this model.

\subsection{It\^o-Schr\"odinger model}
It is proved in \cite{garnier} that in the high frequency and long range regime 
$\lambda_o \ll \ell_c,r_o \ll L$, 
the statistical distribution of the solution of (\ref{eq:para1}--\ref{eq:para2}) can be approximated by that of the 
solution of the It\^o-Schr\"odinger equation 
$$
d u(\bx,z) = \frac{i}{2k}  \Delta_\perp u(\bx,z) \, dz +\frac{i k}{2} u(\bx,z) \circ dB(\bx,z) ,
$$
with initial condition (\ref{eq:para2}).  We wrote this equation in Stratonovich form, and $B(\bx,z)$ is a Brownian field with covariance function 
$$
\EE \big[ B(\bx,z) B(\bx',z') \big] = \ell_z {\cal R}(\bx-\bx') \min\{z,z'\}.
$$

The statistical moments of the beam can be calculated  using It\^o's formula \cite{Ito}.
The first  moment models the coherent (mean) wave and satisfies the damped Schr\"odinger equation
$$
\partial_z \EE [ u(\bx,z) ]=   \frac{i}{2k}  \Delta_\perp \EE[u(\bx,z)]    - \frac{k^2 \sigma^2 \ell_z}{8}
\EE[ u(\bx,z)],
$$
which can be solved explicitly to obtain
\begin{equation}
\EE[u(\bx,z) ]= \int_{\RR^2} u_o(\bx') G\left((\bx,z),(\bx',0)\right) d \bx'  \exp\Big(- \frac{k^2 \sigma^2 \ell_z z}{8}
\Big).
\label{eq:meanu}
\end{equation}
The first factor (the integral)  is the beam in the homogeneous medium propagated using  the  paraxial Green's function~(\ref{eq:GF}). 
The exponential 
decay models the loss of coherence of the beam due to scattering in the random medium.

The second moments of the beam can be described using the mean Wigner transform
\begin{equation}
{\mathcal W} ( \bx,\bk,z) = 
\int_{\RR^2}
\exp \big( - i  \bk \cdot \by  \big)
\EE \Big[ u\Big( \bx+\frac{\by}{2},z \Big)
 \overline{u} \Big(   \bx-\frac{\by}{2},z\Big) \Big]  d \by  ,
 \end{equation}
where the bar denotes complex conjugate. This  satisfies the  radiative transport  equation \cite{garnier}
\begin{align}
 {\partial_z {\mathcal W}(\bx,\bk,z) } + \frac{\bk}{k}
\cdot \nabla_{\bx} {\mathcal W}(\bx,\bk,z)
=\frac{k^2 \ell_z}{4 (2\pi)^2} 
\int_{\RR^2} \hat{\cal R}( \bk - \bk') \nonumber \\
\times \Big[ 
{\mathcal W} ( \bx,  \bk',z   ) 
 - {\mathcal W}( \bx, \bk,z ) 
\Big]  d \bk' ,
\end{align}
which can be solved explicitly
\begin{align}
{\cal W}(\bx,\bk,z) = \frac{1}{(2\pi)^2} \iint_{\RR^2 \times \RR^2}   \hat{\cal W}_o (\bzeta,\by)  
\exp \left[ i \bzeta \cdot \Big(\bx - \bk \frac{z}{k}\Big)\right]  \nonumber \\
\times 
\exp\left[ - 
i \bk \cdot\by+ \frac{k^2 \ell_z}{4} \int_0^z {\cal R}\Big(\by+\bzeta \frac{z'}{k}\Big)-{\cal R}({\bf 0})dz'\right] d\bzeta d\by,
\end{align}
for $\hat{\cal W}_o$  defined in terms of the initial beam profile (\ref{eq:para2}) by
\begin{equation}
\hat{\cal W}_o(\bzeta,\by) = \int_{\RR^2}\exp(- i \bzeta \cdot \bx) u_o\Big(\bx+\frac{\by}{2}\Big)
\overline{u_o}\Big(\bx- \frac{\by}{2}\Big) d\bx.
\end{equation}

\subsection{Weakly diffraction regime}
If the initial radius $r_o$ of the beam and the correlation  radius  $\ell_c$ satisfy the scaling relations 
$
k r_o^2 \gg z$ and   $k \ell_c^2 \gg z,
$
we have a weak diffraction regime,  where the expressions of the first and second moments of the beam simplify to 
\begin{eqnarray}\label{eq:KScoh}
\EE[u(\bx,z) ]=  {u}_o(\bx)\exp\Big(- \frac{k^2 \sigma^2 \ell_z z}{8} \Big),
\end{eqnarray}
and
\begin{align}
\EE \Big[  u\Big(\bx+\frac{\by}{2},z\Big)\overline{u}\Big(\bx- \frac{\by}{2},z\Big)  \Big]
=
 u_o\Big(\bx+\frac{\by}{2}\Big)\overline{u_o}\Big(\bx- \frac{\by}{2}\Big)   \nonumber \\
 \times \exp\left[ \frac{k^2 \ell_z z}{4} \big( {\cal R}(\by )-{\cal R}({\bf 0}) \big) \right] . \label{eq:secondMom}
\end{align}
This corresponds to multiplying the 
initial beam profile $u_o(\bx) $ with a global phase screen.

\section{Demultiplexing and channel efficiency}
\label{sec:3}
We now use the results in section \ref{sec:2} to quantify the recovery of a single mode transmitted through a turbulent random medium. The recovery (demultiplexing) amounts to projecting the received wave field
onto the basis of the transmitted modes and looking at the detected powers.
We compare the efficiencies of the different orthogonal beam families discussed in Section \ref{sec:1}.
We use throughout the setup described in section \ref{sec:illustrate}, where the apertures of the transmitter and receiver arrays are the
same disk ${\cal A}$ of radius $a$.

\subsection{SVD based multiplexing}
\label{sec:SVDMIMO}
For the SVD based MIMO scheme, suppose that  the transmitter array transmits the $j$-th homogeneous input mode $u(\bx,z=0)=u_j^{\rm IN}(\bx)$ for $j \ge 0$,  and the receiver array projects the beam $u(\bx,z=L)$ transmitted through the turbulent medium  onto the homogeneous output modes $u_l^{\rm OUT}(\bx)$, for $l\geq 0$. 
The projection coefficients are defined by
\begin{equation}
\label{def:plj}
p_{l,j} = 
\frac{\big| \int_{{\cal A} }    u(\bx,L) \overline{u_l^{\rm OUT}}(\bx) d\bx  \big|^2}{
 \int_{{\cal A} }  |  u(\bx,L)|^2 d\bx}.
\end{equation}
Since $(u_l^{\rm OUT})_{l \geq 0}$ is a complete orthonormal basis of $L^2( {\cal A} )  $, 
the sum of these non-negative coefficients is $
\sum_{l=0}^\infty p_{l,j}=1 .
$
The mode $u(\bx,z=0)=u_j^{\rm IN}(\bx)$ is well transmitted when $p_{j,j}$ is close to one, so we can call this coefficient the channel efficiency.

In section \ref{sec:numer} we calculate the coefficients (\ref{def:plj}) using the phase screen method described in section \ref{sec:2.1}. We also compare them with the theoretical predictions of the  It\^o-Schr\"odinger model in the weakly diffractive regime, obtained by taking the expectation
in \eqref{def:plj} and using the simple second moment formula (\ref{eq:secondMom}),
\begin{align}
\nonumber
{\cal P}_{l,j} =& \iint_{{\cal A}^2}    u_j^{\rm OUT}(\bx) \overline{u_j^{\rm OUT}}(\bx') \\
&\times
 \overline{u_l^{\rm OUT}}(\bx) u_l^{\rm OUT}(\bx')  {\cal K}(\bx-\bx') d\bx d\bx' ,
\label{eq:theoPjl}
\end{align}
where
\begin{equation}
 {\cal K}(\bx) = \exp \left[ \frac{k^2 \ell_z L}{4} \big( {\cal R}(\bx)-{\cal R}({\bf 0})\big)\right].
 \label{eq:kernK}
\end{equation}
 In particular, the predicted channel efficiency is 
\begin{equation}
{\cal P}_{j,j} = \iint_{{\cal A}^2}   | u_j^{\rm OUT}(\bx)|^2| u_j^{\rm OUT}(\bx')|^2
  {\cal K}(\bx-\bx') d\bx d\bx' .
  \label{eq:MIMOChEf}
\end{equation}

Note that since we assume identical transmitter and receiver apertures, the Rayleigh length calculated with the initial beam profile radius 
(\ref{def:radiusLG}) equals the transmission distance $L$, so diffraction plays a role in the simulations. Nevertheless, the results in section 
\ref{sec:numer}
turn out to be in good agreement with the theoretical prediction estimates (\ref{eq:theoPjl}--\ref{eq:MIMOChEf}).

\subsection{OAM multiplexing}
Let us index the OAM modes by their topological charge \rc{$j \in \mathbb{Z}$} in the phase $\exp(i j \theta)$, which is natural for the Bessel-Gauss beams.
The Laguerre-Gauss beams have a second index, but we already know from section \ref{sec:2} that the significant such modes (in terms of power 
in the aperture) are basically the same as the modes obtained with the SVD approach, discussed above. Therefore, here we focus attention 
on the Bessel-Gauss beams.

When the transmitter array emits the beam $u^{\rm BG}_j(\bx,z=0)$ defined in \eqref{eq:inputBG},  
the receiver array projects the transmitted beam $u(\bx,z=L)$ onto the theoretical profile $u_{l}^{\rm BG}(\bx,z=L)$ given by \eqref{eq:outputBG}. 
This corresponds to defining the  projection coefficients 
\begin{equation}
\label{def:pljBG}
p_{l,j}^{\rm BG} := \frac{\big| \int_{{\cal A} }  u(\bx,L) \overline{u_{l}^{\rm BG}}(\bx,L) d\bx \big|^2}{
\int_{{\cal A} }   |u (\bx,L)|^2 d\bx \int_{{\cal A} }   |u_{l}^{\rm BG}(\bx,L)|^2 d\bx } ,
\end{equation}
where $j$ indexes the initial condition.
Note that  $\left(u_{l}^{\rm BG}(\bx,L) \right)_{l\in \mathbb{Z}}$ is not a complete orthonormal basis of $L^2( {\cal A} )  $, so these coefficients 
do not sum to one,
 $\sum_{l=-\infty}^\infty p_{l,j}^{\rm BG} \ne 1.$ 
We can, however, modify \rc{the definition of the coefficients  to recover this  normalization property} \cite{paterson}. The new  coefficients are 
\begin{equation}
\label{def:pljOAM}
p_{l,j}^{\rm OAM} := \frac{\int_0^a   \big| \int_{0}^{2\pi} u(r,\theta,L) e^{ - i  l  \theta} d\theta \big|^2 rdr }{
2\pi \int_0^a \int_{0}^{2\pi}   |u (r,\theta,L)|^2d\theta rdr}, 
\end{equation}
and they satisfy 
$
\sum_{l=-\infty}^\infty p_{l,j}^{\rm OAM} =1,
$ by Parseval's equality.

In the weakly diffractive regime we have 
\[
u_{j,o}^{\rm BG}(\bx)  \equiv   u_j^{\rm BG}(\bx,z=0) = u_j^{\rm BG}(\bx,z=L), 
\]
and the theoretical predictions of the coefficients (\ref{def:pljBG}) and (\ref{def:pljOAM}) given by the It\^o-Schr\"odinger model are 
\begin{equation}
\label{eq:theoPjlBG}
{\cal P}_{l,j}^{\rm BG} =
\frac{ \iint_{{\cal A}^2}    u_{j,o}^{\rm BG}(\bx) \overline{u_{j,o}^{\rm BG}}(\bx')
 \overline{u_{l,o}^{\rm BG}}(\bx) u_{l,o}^{\rm BG}(\bx')  {\cal K}(\bx-\bx')  d\bx d\bx'}{
\big[ \int_{{\cal A}}   |u_{j,o}^{\rm BG}(\bx)|^2 d\bx \big] \big[ \int_{{\cal A}}   |u_{l,o}^{\rm BG}(\bx)|^2 d\bx \big]  }
\end{equation}
and
\begin{align}
\label{eq:theoPjlOAM}
 {\cal P}_{l,j}^{\rm OAM} =&
\frac{ {\cal N}_{l,j}^{\rm OAM} }{ {\cal D}_{l,j}^{\rm OAM} }, \\
\nonumber
{\cal N}_{l,j}^{\rm OAM} =&
\int_0^a   \iint_{[0,2\pi]^2}   u_{j,o}^{\rm BG}(r,\theta) \overline{u_{j,o}^{\rm BG}}(r,\theta') \\
&\times
 {\cal K}_{{\rm OAM}}(r,\theta-\theta')  e^{ i l (\theta' -\theta)}
  d\theta d\theta' rdr  ,
  \label{eq:theoNjlOAM}
  \\
  {\cal D}_{l,j}^{\rm OAM} =& 2\pi \int_0^a \int_0^{2\pi}   |u_{j,o}^{\rm BG} (r,\theta)|^2d\theta rdr   ,
\end{align}
where ${\cal K}(\bx)$ is defined in \eqref{eq:kernK} and 
\begin{equation}
 {\cal K}_{{\rm OAM}}(r,\theta) = \exp \left[   \frac{k^2 \ell_z L}{4} \big({\cal R}_\nu \big(2 r \big|\sin( \frac{\theta}{2})\big| \big) -{\cal R}_\nu (0)\big)\right].
\end{equation}

\begin{figure}[htbp] 
\begin{center}
\begin{tabular}{c}
    \includegraphics[width=4.3cm]{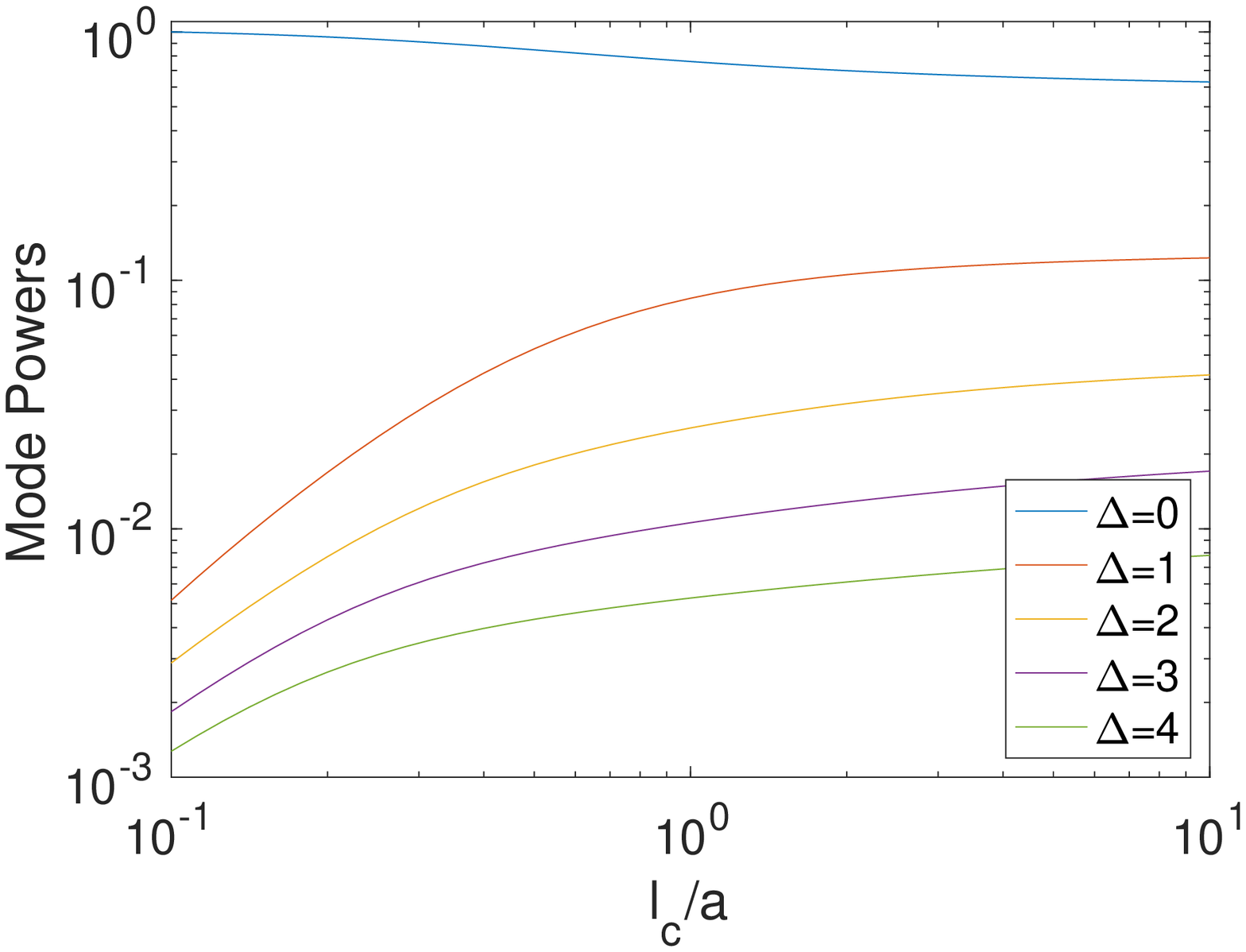}   \includegraphics[width=4.3cm]{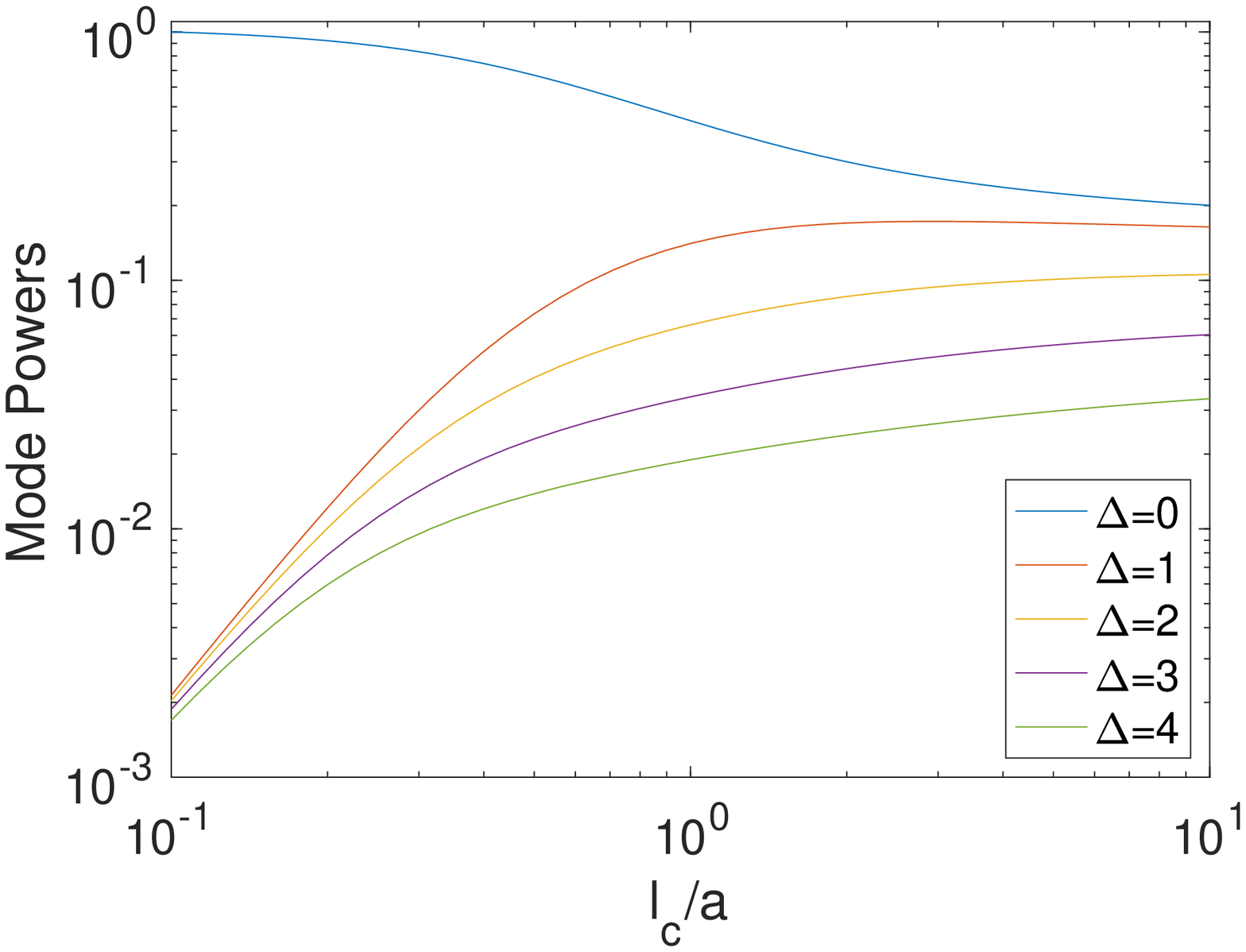} 
    \end{tabular}
\end{center}
\vspace{-0.2in}
    \caption{Theoretical prediction \eqref{eq:theoPjlOAM} of the projection power for $j=1$ (left plot) and $j=9$ (right plot) as a function of 
    $\ell_c/a$. The different curves are for the values of   $\Delta = |l-j|$ shown in the legend. }
    \label{fig:OAM}
 \end{figure}

For illustration, we plot in  Figure \ref{fig:OAM} the coefficients ${\cal P}_{l,j}^{\rm OAM}$  as a function of $\ell_c/a$ for $j=1$  (left plot) and $j=9$ (right plot). 
We consider various values of $\Delta = |l-j|$  and note that the sign of $l-j$  does not affect ${\cal P}_{l,j}^{\rm OAM}$ 
\rc{in view of (\ref{eq:inputBG}) and (\ref{eq:theoNjlOAM})}.
We use
$\sigma^2 k^2 \ell_z L =8,~r_o/a=1,~\beta a= 6,~\nu= 5/6$, 
\rc{so that the mean field is strongly damped,  but not completely vanished (see (\ref{eq:KScoh})), 
there is a strong interaction between the aperture and the initial radius, and 
the variations in the source Bessel function are captured within the  initial  radius}.    
Similar results were obtained in  \cite{paterson,tyler} based on a formal model for the effect of the turbulence in the form of a phase screen. 
Here we have put this model in the mathematical framework of beam propagation in random media, which makes explicit the scaling regime where
it  is valid, and we clarified the link between the  phase screen parameters and those of the model for the physical medium.  
The figure shows that  the cross talk between the modes becomes noticeable in a regime corresponding to $\ell_c \gtrsim a$. The numerical simulations in the next section are  in this regime. 

\subsection{Numerical simulations}
\label{sec:numer}
We now present numerical results obtained with the phase screen method described in section \ref{sec:2.1}, for the setup  in section 
\ref{sec:illustrate}, and the Kolmogorov-type 
model of the covariance obtained from  definition (\ref{def:materncov}). The hyperparameters in this model are  $\nu = 5/6$, $\ell_c = 20$cm (i.e., $L_0 = 37$cm) and we consider three values of $C_n^2$ corresponding to a homogeneous medium $(C_n^2 = 0)$, weak turbulence ($C_n^2 = 10^{-14})$ and stronger turbulence ($C_n^2 = 4 \cdot 10^{-14}$). These choices are similar to those in \cite{doster,paterson}, except for the outer scale $L_0$, which is smaller in our simulations. This does not have a big effect because the radius of the beams  is smaller than the radius $a = 5$cm of the apertures and therefore smaller than  $L_0$. The wavelength is $\lambda = 850$nm and the transmission distance is $L = 1$km.

The SVD based multiplexing is carried out using the SVD of the transfer matrix in the synthetic homogeneous medium. The transmitter and receiver 
aperture ${\cal A}$ is as in the left plot in Figure \ref{fig:circularplanararray}, so in the simulations
 we shift the beam axes to pass through the center \rc{$(5,5)$}cm of ${\cal A}$.

\begin{figure}[th]
 \begin{center}
  \begin{tabular}{c}
\includegraphics[width=2.85cm]{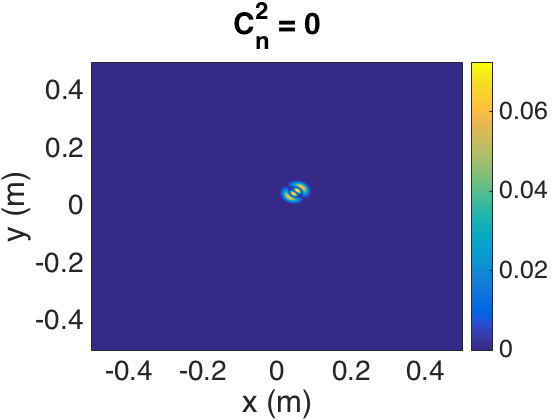}  
\includegraphics[width=2.85cm]{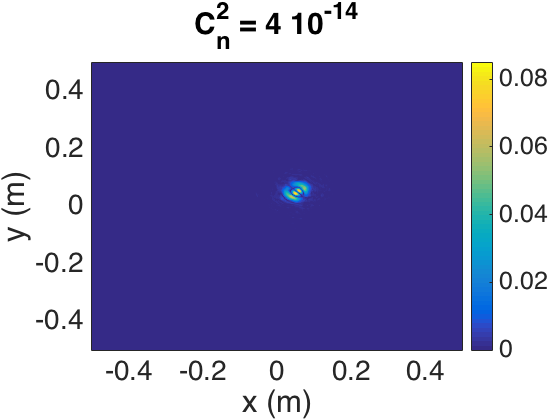}  
\includegraphics[width=2.85cm]{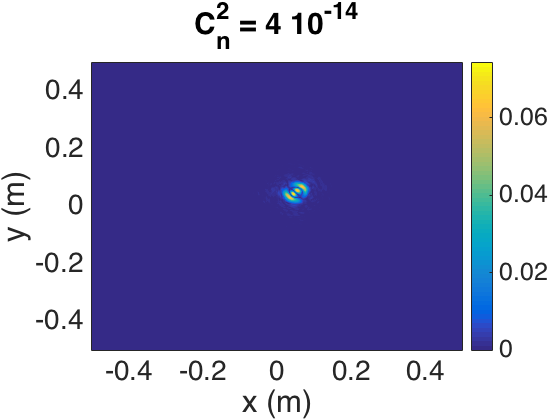}  \\
\includegraphics[width=2.85cm]{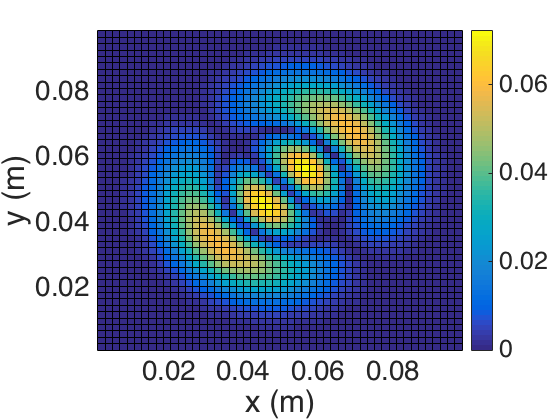}  
\includegraphics[width=2.85cm]{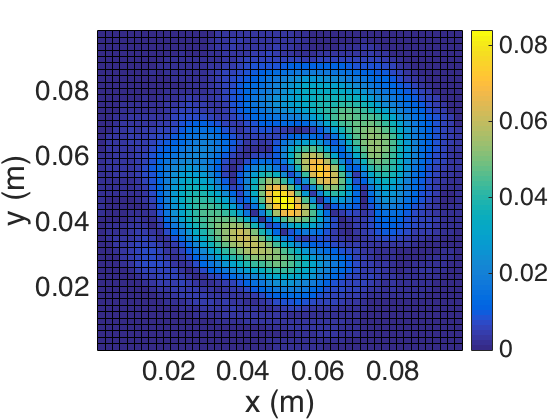} 
\includegraphics[width=2.85cm]{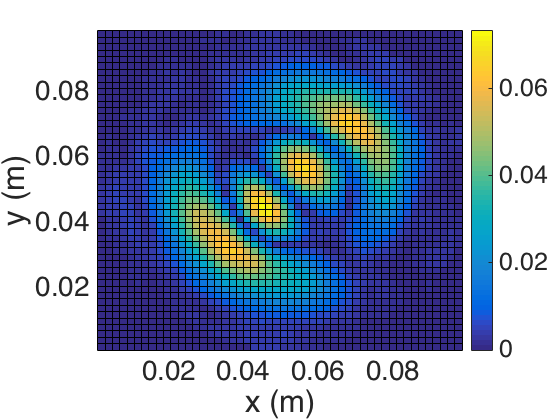}  \\
\includegraphics[width=2.85cm]{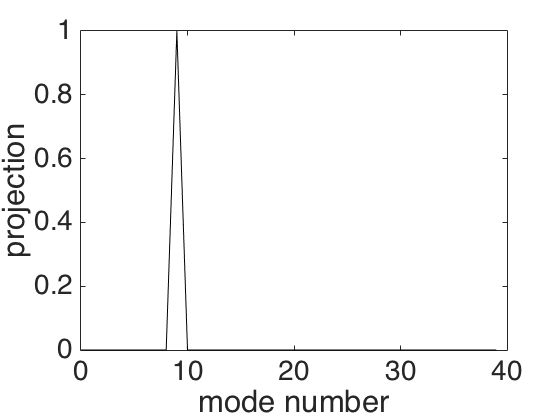} 
\includegraphics[width=2.85cm]{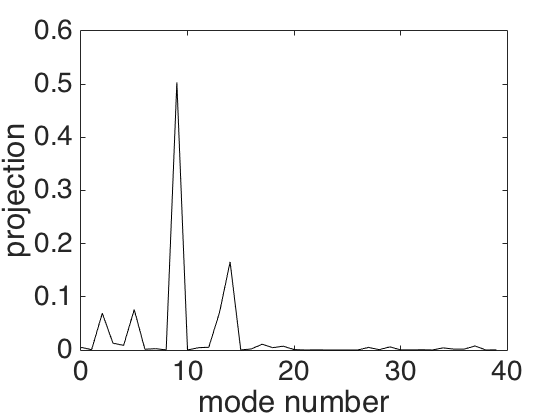} 
\includegraphics[width=2.85cm]{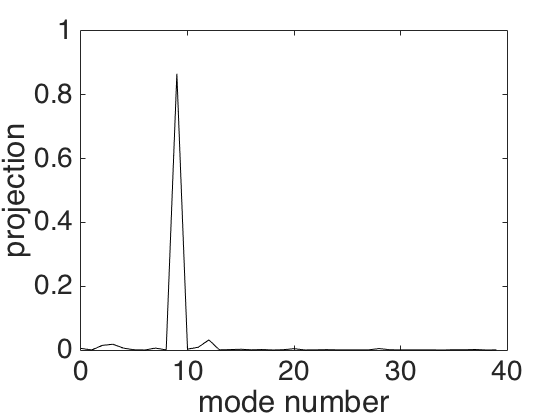} 
\end{tabular}
\end{center}
\vspace{-0.2in}\caption{Here the input beam is the $j=9$th  mode obtained from the SVD. 
The top two rows show the transmitted field moduli at the receiver array. 
The second row is a zoom of the first, more exactly the values of the modulus of the field recorded by the receiver array. 
The axes are the coordinates in the cross-range plane, in meters.
The bottom  row plots the projection coefficients $p_{l,9}$ for $l=0,\ldots,39$. The left column is for the homogeneous transmission medium. The middle and right columns are for two realizations 
of the random medium with turbulence level $C_n^2 = 4 \cdot 10^{-14}$. }
\label{fig1a}
\end{figure}

In the top two rows of Figure \ref{fig1a} we display the modulus of the beam at the receiver array, due to the initial profile given by 
the $j=9$-th  input SVD mode in Figure \ref{fig:eigen}.  The results are obtained  in two realizations of the turbulent random medium, for the 
stronger turbulence ($C_n^2 = 4 \cdot 10^{-14}$). We also display the projection coefficients \eqref{def:plj}. As expected, the channel efficiency 
is perfect ($p_{j,j}=1$) in the homogeneous medium, and it deteriorates in the turbulent medium, due to 
mode mixing, and the result is dependent on the realization of the medium.

\begin{figure}[th]
 \begin{center}
  \begin{tabular}{c}
\includegraphics[width=2.85cm]{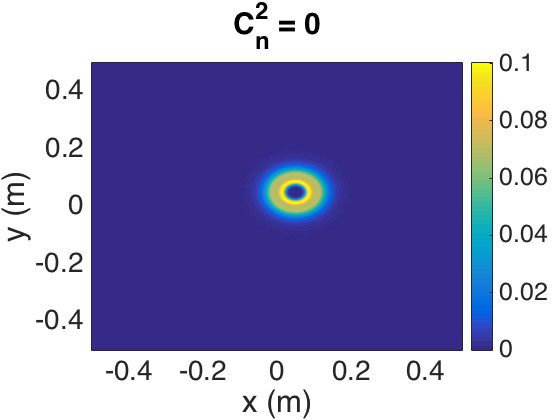}  
\includegraphics[width=2.85cm]{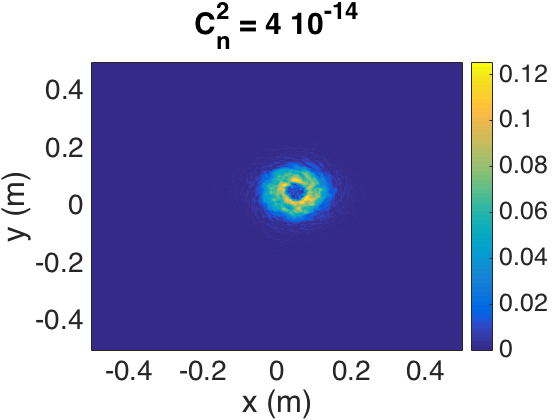}  
\includegraphics[width=2.85cm]{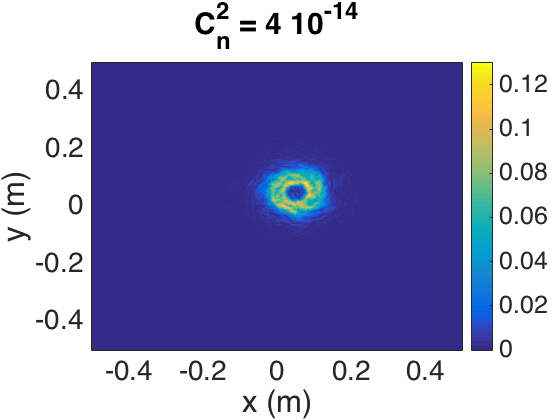}  \\
\includegraphics[width=2.85cm]{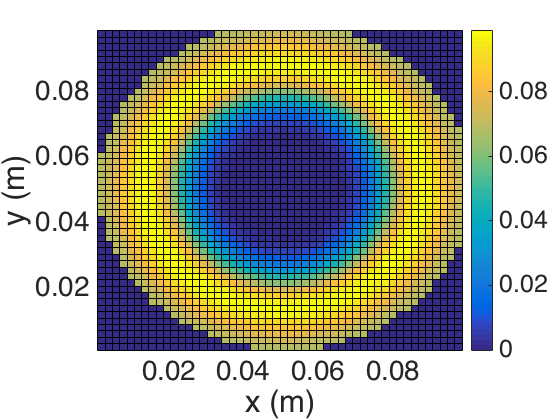}  
\includegraphics[width=2.85cm]{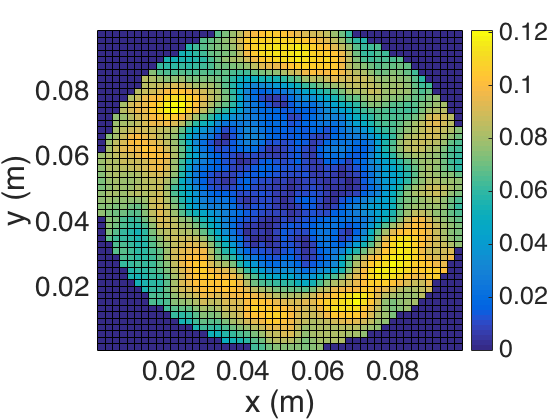}  
\includegraphics[width=2.85cm]{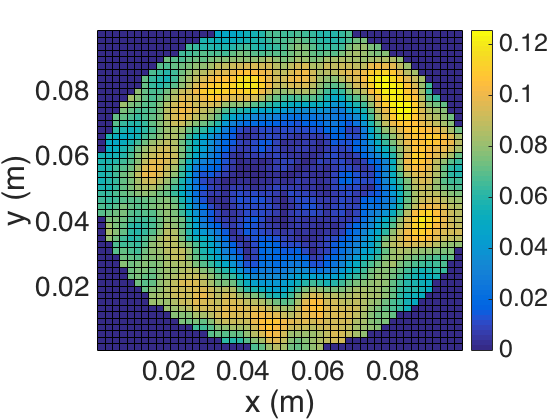}  \\
\includegraphics[width=2.85cm]{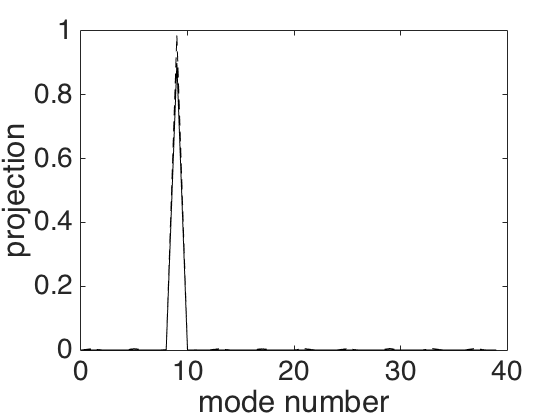} 
\includegraphics[width=2.85cm]{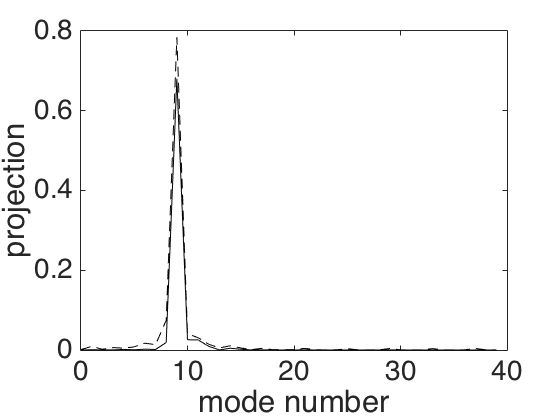} 
\includegraphics[width=2.85cm]{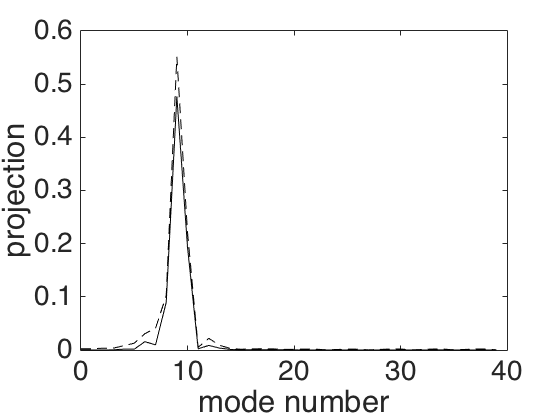} 
\end{tabular}
\end{center}
\vspace{-0.2in}
\caption{Here the input beam \rc{is} the $j=9$th input Bessel-Gauss mode with radius $r_o$ given by (\ref{def:radiusLG}). 
The top two rows show the transmitted field moduli at the receiver array. 
The second row is a zoom of the first. The axes are the coordinates in the cross-range plane, in meters.
The bottom  row plots the projection coefficients $p_{l,9}^{\rm BG}$ (solid line) and $p_{l,9}^{\rm OAM}$ (dashed line) for $l=0,\ldots,39$. 
The left column is for the homogeneous transmission medium. The middle and right columns are for two realizations 
of the random medium with turbulence level $C_n^2 = 4 \cdot 10^{-14}$. }
\label{fig1b}
\end{figure}
\begin{figure}[th]
\begin{center}
\begin{tabular}{c}
\includegraphics[width=4.35cm]{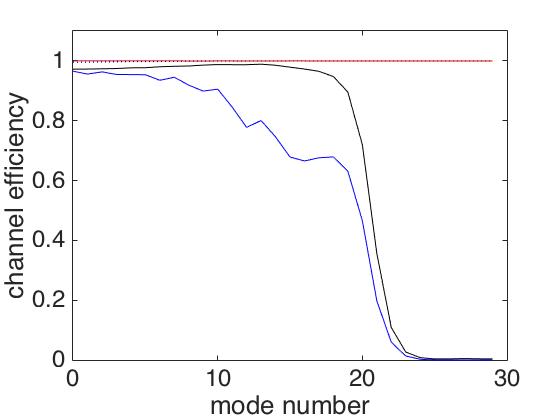} 
\end{tabular}
\end{center}
\vspace{-0.2in}
\caption{Channel efficiencies in the homogeneous medium:  $p_{j,j}^{\rm BG}$ (solid blue) and $p_{j,j}^{\rm OAM}$ (solid black) for the Bessel-Gauss modes and $p_{j,j}$ (solid red) for the SVD modes.}
\label{fig:channelefficiency0}
\end{figure}

Figure \ref{fig1b} is the analogue of Figure \ref{fig1a}, except that the input beam is the $9-$th Bessel-Gauss mode.
The main difference between Figures \ref{fig1a} and \ref{fig1b} 
is that the power delivered by the Bessel-Gauss beam is mostly on the edges of the aperture, whereas for the SVD mode the power
is well contained inside the aperture. This plays a role for higher mode numbers, because the Bessel-Gauss modes do not take the 
finite aperture into account and they deliver less and less power within the receiver array. See  Figure \ref{fig:channelefficiency0} for an illustration 
of this effect in the homogeneous medium.

\begin{figure}[th]
 \begin{center}
  \begin{tabular}{c}
\includegraphics[width=4.35cm]{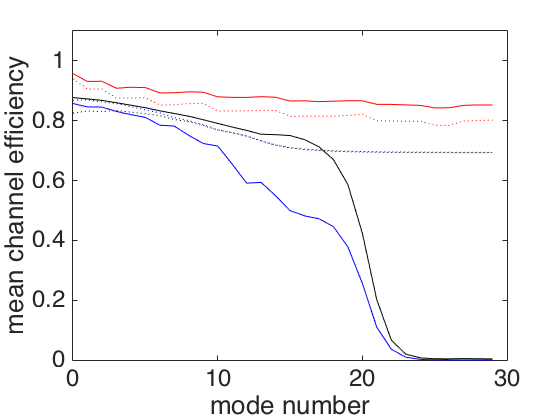} 
\includegraphics[width=4.35cm]{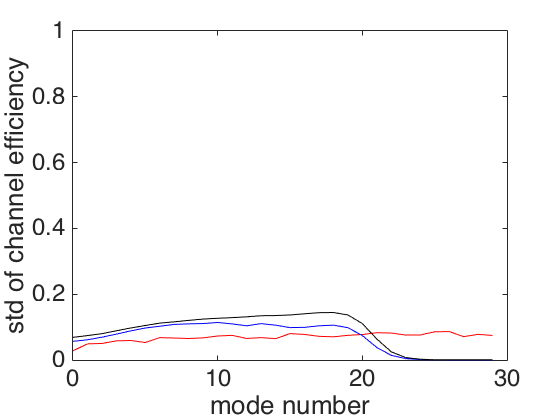} \\
\includegraphics[width=4.35cm]{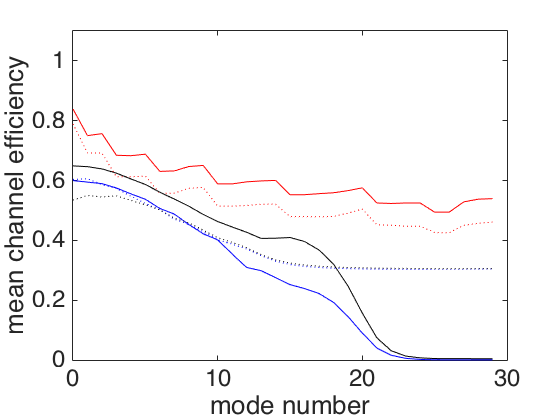} 
\includegraphics[width=4.35cm]{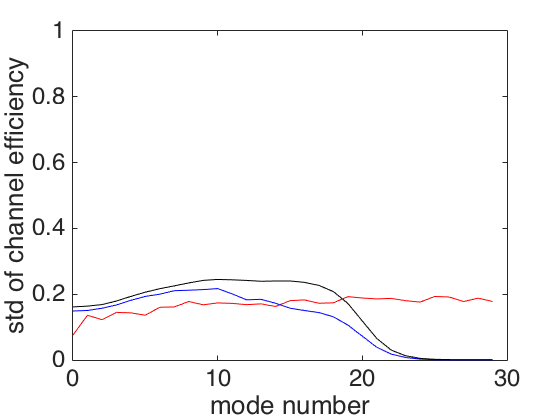} 
\end{tabular}
\end{center}
\vspace{-0.2in}
\caption{Left: Mean channel efficiencies $p_{j,j}^{\rm BG}$ (solid blue) and $p_{j,j}^{\rm OAM}$ (solid black) for the Bessel-Gauss modes and $p_{j,j}$ for the homogeneous SVD modes (solid red). Right: Standard deviations of the channel efficiencies. 
Here $\sigma=3 \, 10^{-9}$, $\ell_c =20$~cm (i.e. $C_n^2 = 4\,10^{-14}$, $L_0=37$~cm),  $a=5$~cm, $\lambda_0=850$~nm, $L=1000$~m.
The dotted lines stand for the theoretical formulas (\ref{eq:theoPjlBG})  (dotted blue), 
 (\ref{eq:theoPjlOAM})  (dotted black), 
(\ref{eq:theoPjl}) (dotted red). Top row weak turbulence ($C_n^2 = 10^{-14}$) and bottom row stronger turbulence ($C_n^2 = 4 \cdot 10^{-14}$).}
\label{fig:channelefficiency15}
\end{figure}

The plots in Figures \ref{fig1a} and \ref{fig1b} show that the channel efficiency varies from one realization of the random medium  to another. Therefore, we display in 
Figure \ref{fig:channelefficiency15} the mean channel efficiency obtained by averaging over 100 realizations of the random medium, and its standard deviation. The solid lines in these figures show the performance of Bessel-Gauss (black and blue lines) and SVD modes (red lines). 
We also plot with the dotted  lines the theoretical predictions given by the It\^{o}-Schr\"{o}dinger model in the weakly diffractive regime.
It appears that the low-order Bessel-Gauss modes are approximately as good as the low-order SVD modes. However, the mean channel efficiency of the Bessel-Gauss modes decreases much faster with the mode number. The channel efficiencies of the SVD modes also have 
smaller standard deviation. 

The comparison between the red solid and dotted lines in the left plots of Figure \ref{fig:channelefficiency15}  shows a good quantitative agreement between formula (\ref{eq:theoPjl}) and the numerical simulations. This is because the profiles of the transmitted SVD modes 
are well captured by the receiver array and the predictions of the It\^o-Schr\"odinger model in the weakly diffractive regime are reliable.
When comparing the solid and dotted black and blue lines, we observe only qualitative agreements between formulas (\ref{eq:theoPjlBG}-\ref{eq:theoPjlOAM}) and the numerical simulations. This is because the profiles of the transmitted Bessel-Gauss modes are poorly
captured by the receiver array (the  modes are concentrated on a thin annulus which diffracts) and the predictions of the It\^o-Schr\"odinger model in the weakly diffractive regime are then not reliable.

\section{Summary}
\label{sec:summary}
We introduced a mathematical framework for studying MIMO and OAM multiplexing for free-space optical communications between a transmitter and receiver array, using laser beams. The study takes into account the finite apertures of the arrays and the scattering of a turbulent transmission medium. For the commonly used circular apertures, we connected the two multiplexing approaches using the theory of prolate spheroidal functions.
Explicitly, we showed that in regimes with a large number of significant singular values of the transfer matrix (i.e., many modes available for multiplexing), the MIMO approach is the same as the OAM approach for Laguerre-Gauss vortex beams, provided these have a well callibrated 
initial radius that depends on the wavelength, the distance of propagation and the ratio of the radii of the transmitter and receiver apertures. 
These communication modes are superior to other vortex beams, for example Bessel-Gauss, which do not take the finite aperture effect into account. 

We used the theory of beam propagation in random media to put the phase screen numerical propagation method in a mathematical
framework and to clarify the dependence of the phase screen parameters on the Kolmogorov-type model of turbulence. The theory 
gives theoretical estimates of the communication channel efficiency, which are compared with numerical results obtained with the phase screen method. The results demonstrate  the superior performance of the SVD based multiplexing/demultiplexing approach for communication through a turbulent medium.

\section*{Acknowledgements}
This research is supported in part by AFOSR grants FA9550-18-1-0131 and FA9550-18-1-0217,  by ONR grant N00014-17-1-2057
and NSF  grant 1616954. The work of the second author was also partially supported by the French ANR under Grant No. ANR-19-CE46-0007 (project ICCI).


\bibliography{bibli}\bibliographystyle{siam}

\end{document}